\documentclass[twocolumn,tighten]{aastex631}

\usepackage{graphicx}
\usepackage{amsmath}	
\usepackage{amssymb}	
\usepackage{bbold}	    
\usepackage{amsfonts}
\usepackage{mathrsfs}
\usepackage{nicefrac}

\usepackage{amsmath}
\usepackage{amssymb}
\usepackage{upgreek}
\usepackage{bbm}
\usepackage{dsfont}
\usepackage{sansmath}
\usepackage{mathrsfs}
\usepackage{yfonts}
\usepackage{euscript}

\def\d{{\rm d}}
\def\D{{\rm D}}
\def\del{\partial}
\def\bfnabla{\mbox{\boldmath{$\nabla$}}}

\def\R{{\mathbbm{R}}}
\def\calA{{\cal A}}

\def\calH{{\cal H}}
\def\calQdot{\dot{\cal Q}}

\def\bfzeta{\mbox{\boldmath{$\zeta$}}}
\def\bfeta{\mbox{\boldmath{$\eta$}}}
\def\bfpsi{\mbox{\boldmath{$\psi$}}}

\def\bfCalD{\mbox{\boldmath{${\cal D}$}}}

\def\bfiV{\boldsymbol{V}}
\def\bfiu{\boldsymbol{u}}
\def\bfiv{\boldsymbol{v}}
\def\bfix{\boldsymbol{x}}
\def\bfif{\boldsymbol{f}}

\def\scrD{\mathscr{D}}

\def\scrT{\mathscr{T}}
\def\scrK{\mathscr{K}}
\def\scrR{\mathscr{R}}
\def\scrQdot{\dot{\mathscr{Q}}}

\def\gsim{\;\rlap{\lower 2.5pt\hbox{$\sim$}}\raise 1.5pt\hbox{$>$}\;}
\def\lsim{\;\rlap{\lower 2.5pt\hbox{$\sim$}}\raise 1.5pt\hbox{$<$}\;}
\def\del{{\partial}}

\def\bfF{\mbox{\boldmath{$F$}}}

\def\bfzeta{\mbox{\boldmath{$\zeta$}}}

\def\beq{\begin{equation}}
\def\eeq{\end{equation}}

\usepackage{color}
\definecolor{com_red}{rgb}{.8,.2,0.1}
\definecolor{com_burg}{rgb}{.45,.05,0.15}
\definecolor{com_turq}{rgb}{.00,.35,0.65}
\definecolor{ins_green}{rgb}{.1,.5,0.1}

\allowdisplaybreaks[1]

\shorttitle{The Diabatic Equivalent-Barotropic Model}
\shortauthors{Wei, Kafle, \& Cho}

\graphicspath{{./}{figs/}}

\begin{document}

\title{Atmospheric Circulation of Close-In Extrasolar Giant Planets:\\ 
The Diabatic Equivalent-Barotropic Model}

\author[0009-0001-3400-6940]{Songyuan Wei}
\affiliation{Martin A. Fisher School of Physics, Brandeis University, 
415 South Street, Waltham, MA 02453, USA}
\correspondingauthor{Songyuan Wei}

\author[0000-0003-2649-7978]{Jagat Kafle}
\affiliation{Martin A. Fisher School of Physics, Brandeis University, 
415 South Street, Waltham, MA 02453, USA}

\author[0000-0002-4525-5651]{James Y-K. Cho}
\affiliation{Martin A. Fisher School of Physics, Brandeis University, 
415 South Street, Waltham, MA 02453, USA}

\email{Emails:\ songyuanwei@brandeis.edu, jagatkafle@brandeis.edu,\\ jamescho@brandeis.edu}

\begin{abstract}
We extend the description of equivalent-barotropic equations for 
exoplanets to the diabatic case---that is, with explicit 
heating and/or cooling representation, rather than with a stationary 
deflection of the bottom bounding surface. 
In the diabatic case, the equation for potential temperature (or 
entropy) is directly forced and cannot be decoupled from the equations 
for momentum and nonlinear pressure, the mass-like variable; 
and, the isentropic surfaces do not remain coincident with material 
surfaces. 
Here the formalism is presented for an atmosphere with the Lamb 
vertical structure, as the formalism is substantially simplified 
under the structure.  
The equations presented set the stage for accurate global simulations 
which permit small-scale vortices, gravity waves, and fronts observed 
in current three-dimensional global simulations to be studied in 
detail.
\end{abstract}


\section{Introduction}\label{sec:intro}

The dynamical essence of a planetary atmosphere can often be captured 
and studied with a thin layer representation. 
Such an atmosphere would be stably stratified and typically exhibit 
strongly barotropic (vertically aligned) flows over some depth.
This applies to Earth's stratosphere and appears to be the case for 
giant planets, including giant exoplanets 
\citep[e.g.,][]{ChoPol96b,SkinCho22}.
A good understanding of dynamics is crucial for interpreting and 
planning observations from JWST \citep[e.g.,][]{Gardneretal06} and 
Ariel \citep[e.g.,][]{Tinettietal21} missions.
Close-in extrasolar giant planets (CEGPs), whether close-in over the 
entire orbit or only part of the orbit, are also important for theory 
since the thermal forcing expected for them present an idealized 
configuration for instructive studies.

In this letter, as in \cite{Choetal03,Choetal08}, we use the 
``equivalent-barotropic'' formulation of the primitive 
equations \citep[PEs; see, e.g.,][]{Salby96}.
The equivalent-barotropic equations (EBEs) facilitate a clear 
understanding of certain physical mechanisms in isolation (e.g., 
the effects of vortices, waves, and heating and cooling on 
columnar motions) while bypassing the issue of current lack of 
information on the vertical thermal structure and distribution 
of radiatively and/or chemically active species.
By construction, like the more familiar shallow-water equations
\citep[SWEs; e.g.,][]{Pedl87}, the EBEs cannot formally address 
baroclinic (vertically slanted) flows.
However, unlike the SWEs, the EBEs are valid even when the 
density $\rho$ or temperature $T$ of the modeled layer is not 
uniform \citep[see, e.g.,][]{Choetal08}. 
Hence, while the SWEs are more appropriate for solar system 
giant planets, given their nearly uniform temperatures
at the cloud decks \citep[e.g.,][]{ChoPol96b,ChoPol96a}, 
the SWEs are not the proper equations for studying CEGPs,
in general.
For atmospheres which exhibit strong barotropicity and lateral 
variation in $\rho$ or $T$, including those on CEGPs 
\citep{ThraCho10,SkinCho21}, the EBEs are appropriate.
Here we focus on the diabatic (i.e., with explicit thermal 
forcing) extension of the adiabatic EBEs, employed 
by \citet{Choetal03,Choetal08}.

Currently, a primary focus of exoplanet characterization studies 
is the large-scale atmospheric structure, particularly the 
temperature and flux distributions resulting from the atmospheric 
motions \citep[e.g.,][]{Kempetal23,Changeatetal24}.
However, the flows are not well-resolved---even at the large scale, 
in many cases---and therefore poorly modeled in current numerical 
simulations \citep[see, e.g.,][]{Choetal03,Choetal15,SkinCho22}.
In most simulations, small-scale gravity waves are not resolved at 
all; and, these waves have a significant influence on the 
large-scale flow via momentum and heat transport as well as mixing 
of active species \citep[e.g.,][]{WatCho10}. 
The sources of gravity waves are 
varied \citep[e.g.,][]{Hami97,FritAlex03}, and quantification of 
their influence on the large-scale motion and temperature 
distribution has been a long-standing problem in atmospheric 
studies \citep[see, e.g.,][and references therein]{Frit84,Lind90, Hami97,FritAlex03,Achaetal24}; 
see also \citet{Choetal21} and \citet{SkinCho24} for discussions 
in the exoplanet context.

It is important to emphasize that the currently observable regions 
of CEGP atmospheres are forced not only by stellar 
irradiation---but also by vortices and waves (both of which have 
long-range as well as long-duration influences).
The vortices and waves actively control the deviation of the 
temperature from the equilibrium distribution established by the 
irradiation.
Therefore, they must be represented accurately in simulations. 
However, accurate representation in three-dimensional (3D) 
calculations are extremely difficult at 
present \citep{Choetal15,SkinCho21,SkinCho24}. 
This can be alleviated by vertically integrating the equations 
typically solved in simulations (the PEs) and using an accurate 
method 
to solve the resulting reduced set of equations (e.g., the EBEs). 
In doing so, some of the elements which are poorly represented can 
begin to be addressed.
There are other justifications for focusing on the EBEs for CEGPs, 
and these are discussed in \citet{Choetal08}. 
Much of the discussion here follows those in \citet{Char49}, 
\citet{Salby89}, and \citet{Choetal08}, and the reader is directed 
to those works for more details as well as derivations.
In this letter, we mainly present the diabatic EBEs in useful 
forms and highlight several salient features.

\section{Diabatic Equations}\label{sec:equations}

The PEs, which govern the large-scale 3D dynamics of the 
atmosphere, are the Navier--Stokes equations for a compressible 
fluid under differential rotation and hydrostatic balance (i.e., 
stable vertical stratification).
Under the adiabatic condition, if an isentrope (a potential 
temperature or entropy surface $\theta$), is initially coincident 
with a material surface, it remains so.
However, under the diabatic condition, $\theta$ is not materially 
conserved and the two surfaces diverge following the flow:
the material surface crosses into another isentrope.
This property offers an advantage in interpreting the equations, 
and their governing dynamics, when formulated in isentropic 
coordinates---i.e., $(\bfix, z, t) \rightarrow 
(\bfix, \theta(\bfix, z, t), t)$, where $\bfix \in \R^2$ and $z$ 
is the height.  
In isentropic coordinates, the PEs read \citep[e.g.,][]{Salby96}:
\begin{subequations}\label{eqn:pe}
    \begin{eqnarray}
        \frac{\scrD \bfiv}{\scrD t} 
        &\ =\ &\,   -\bfnabla \Psi\, -\, \bfif\!\times\!\bfiv\, 
        +\, \bfCalD_{\bfiv}  \qquad \qquad \\
        \frac{\del\Psi}{\del \theta} &\ =\ &\,  \Pi \\
        \frac{\scrD \beta }{\scrD t}
        &\ =\ & -\, \beta\!\left(\bfnabla\cdot\bfiv + 
        \frac{\del\varpi}{\del \theta}\right)\\
        \frac{\scrD \theta}{\scrD t}  &\ =\ & 
        \frac{1}{\Pi}\,\scrQdot.
    \end{eqnarray}
\end{subequations}
Here
$\scrD / \scrD t\, =\, D / Dt + \varpi \del / \del\theta$, where
$D / Dt\, =\, \del / \del t + \bfiv\cdot\bfnabla$ with $\bfnabla$ 
the two-dimensional (2D) gradient operator on an isentrope;
$\bfiu = (\bfiv, \varpi)$ with $\bfiv = \bfiv(\bfix, \theta, t)$ 
the horizontal velocity and $\varpi \equiv \scrD \theta/ \scrD t$ 
the vertical velocity such that $\del \theta / \del z > 0$; 
$\Psi(\bfix, \theta, t) = c_p T + g z_\theta$ is the Montgomery streamfunction, where $c_p$ is the specific heat at constant $p$, 
$T(\bfix, \theta, t)$ is the temperature, $g$ is the gravity, 
and $z_\theta(\bfix, \theta, t)$ is the isentrope elevation;
$\bfif = \bfif(\bfix)$ is the Coriolis parameter, oriented in 
the vertical direction, with $f = |\bfif|$;
$\bfCalD_{\bfiv}(\bfix, \theta, t)$ is the momentum dissipation; 
$\Pi(\bfix, \theta,t) \equiv c_p\,(p/p_{00})^\scrK\! = 
c_p\, (T / \theta)$ is the Exner function, where $p_{00}$ is 
a reference $p$ (a constant), $\scrK \equiv \scrR / c_p$ is the 
Poisson constant with $\scrR = c_v - c_p$  the specific gas 
constant and $c_v$ the specific heat at constant volume; 
$\beta \equiv \del p / \del \theta$; and, 
$\scrQdot(\bfix, \theta,t)$ is the diabatic forcing.

The atmosphere is, in general, baroclinic: two sets of 
thermodynamic surfaces (e.g., $p$ and $\theta$) are independent.
But, there is a special class of baroclinic stratification, the 
equivalent-barotropic stratification, which permits a physically 
valid 2D reduction of the PEs for a fluid layer with lateral 
density or temperature inhomogeneity \citep[e.g.,][]{Salby89}.
Under the equivalent-barotropic stratification, two sets of 
thermodynamic surfaces are not entirely independent: they share a
common horizontal structure.
Due to hydrostatic balance, the pressure gradient force---hence 
the horizontal velocity---are parallel at all heights, \citep{Char49,EliaKlei57}.
In this case, the baroclinic field is separable into a barotropic 
field and a vertical structure function $\calA$.
For example, $\bfiv$ can be separated into
\begin{eqnarray}\label{eqn:separation}
    \bfiv(\bfix, \theta, t)\, =\, 
    \calA(\theta)\,\hat{\bfiv}(\bfix, t),
\end{eqnarray}
where $\calA$ is a scalar function \citep{Char49}.
The separation can be formally effected by a ``barotropic 
transformation'',
\begin{eqnarray}\label{eqn:transform}
       \hat{\bfiv}(\bfix, t)\, =\, -\frac{1}{p_0} 
       \int_{\theta_0 = \theta(\bfix, p_0, t)}^\infty 
       \bfiv(\bfix, p, t) 
       \left(\frac{\del p}{\del \theta}\right) d\theta, \quad
\end{eqnarray}
where the integration is from the bottom of the fluid layer to the 
top of the fluid layer (NB., throughout this letter, a 0-subscript 
on a variable always denotes the variable evaluated at the bottom 
bounding surface; we now also drop the hat on the 
equivalent-barotropic variables, as the equivalent-barotropicity 
is clear from the context).
Equations~(\ref{eqn:separation}) and (\ref{eqn:transform}) form an 
inverse transform pair with the normalization,
\begin{eqnarray}
       -\frac{1}{p_0} 
       \int_{\theta_0}^\infty \calA(\theta) 
       \left(\frac{\del p}{\del \theta}\right) d\theta = 1.
\end{eqnarray}
Therefore, $\calA_0 > 1$ for an equivalent-barotropic structure that 
decays vertically and $\calA_0 < 1$ for a structure that grows 
vertically; if the structure is barotropic, $\calA_0 = 1$.

The EBEs govern the dynamics of a semi-infinite gas layer,  
bounded below by a material surface. 
The bounding surface deforms according to the local $T$ change on 
that surface.
Under diabatic condition, the EBEs in $\theta$-coordinates 
read \citep[e.g.,][]{Salby89,Choetal08}:
\begin{subequations}\label{eqn:debe}
    \begin{eqnarray}
        \frac{D\bfiv}{D t}\ &\ =\ & -\bfnabla \phi_\ast 
        - \bfif\!\times\!\bfiv\, +\, \phi\,\bfnabla(\ln\theta_0) 
        \nonumber \\ 
        & & +\ \frac{1}{\calA_0 (1 + \scrK)
        \theta_{00}}\!\left(\frac{\theta_0}{\phi}\right)
        \bfiv\,\calQdot\, +\, \bfCalD_{\bfiv} \\
        \frac{D \phi}{D t}\ &\ =\ &
        -\scrK \phi\,\bfnabla\!\cdot\!\bfiv\, +\, (1 - \calA_0)\, 
        \phi \bfiv\!\cdot\!\bfnabla(\ln\theta_0) 
        \qquad \quad \nonumber \\
        & & + \frac{1}{\calA_0^{\,\prime}\,\theta_{00}}\,
        \calQdot  \\
        \frac{D\theta_0}{D t} &\ =\ & 
        (1 - \calA_0)\,\bfiv\!\cdot\!\bfnabla\theta_0\, +\, 
        \frac{1}{\calA_0^{\,\prime}\theta_{00}}\!
        \left(\frac{\theta_0}{\phi}\right)\calQdot.
    \end{eqnarray}
\end{subequations}
Here 
$\bfiv = \bfiv(\bfix, t)$ is the equivalent-barotropic velocity;
$\phi_\ast(\bfix,t) \equiv \phi + \phi_b$, where 
$\phi(\bfix,t) = \theta_0 \Pi_0 / \calA_0$ and
$\phi_b(\bfix,t) = g z_0 / \calA_0$ with $z_0(\bfix,t)$ the 
prescribed elevation of the bottom surface (in this letter, 
$\{\scrR, c_p, c_v, \gamma, g, \Gamma\}$, where 
$\gamma \equiv c_p / c_v = 1 / (1 + \scrK)$ is the adiabatic 
index, and $\Gamma \equiv g / c_p$ is the adiabatic lapse rate, 
are all taken to be a constant);
$\theta_{00}$ is a reference $\theta$ (a constant);
$\calA_0^{\,\prime} \equiv (\d\calA\, /\, \d\theta)\,|_{\theta_0}$; 
and, $\bfCalD_{\bfiv}(\bfix,t)$ and $\calQdot(\bfix,t)$ represent 
equivalent-barotropic momentum dissipation and diabatic forcing, 
respectively. 
Note that $\phi = c_p T_0 / \calA_0 = 
g \calH_{p_0} / (\calA_0 \scrK)$, where 
$\calH_{p_0}(\bfix, t) \equiv \scrR T_0 / g$ is the pressure scale 
height at the bottom bounding surface.
Thus, $\phi$ measures changes of temperature along that surface: 
$\phi$ is closely related to the local pressure scale 
height---as well as the local potential temperature scale height,
$\calH_{\theta_0}(\bfix, t) \equiv 
(\del\theta / \del z |_{0})^{-1} = \calH_{p_0} / \scrK $.
Hence, it can be readily seen from Equations~(\ref{eqn:debe}) that, 
in the EBEs, heating and cooling forces the flow through the 
deflection of the bottom bounding surface, which advects the 
temperature.
The advected temperature in turn drives the flow when gradients 
form on the surface.

Equations~(\ref{eqn:debe}) also admit an important conservation law 
for a potential vorticity~$q$ \citep{Salby89}: 
\begin{eqnarray}\label{eqn:q}
    \frac{D q}{D t} &\ =\ &  \frac{1}{\phi^{\nicefrac{1}{\scrK}}}
    \Big|\bfnabla\phi\!\times\!\bfnabla(\ln\theta_0)\Big| 
    -\, \frac{1 - \calA_0}{\scrK}\,
    q\bfiv\!\cdot\!\bfnabla(\ln\theta_0)\nonumber \quad \\  
    & & +\, \frac{1}{\phi^{\nicefrac{1}{\scrK}}} 
    \Big|\bfnabla\! \times\! \bfCalD_{\bfiv} \Big| \quad \nonumber \\
    & & +\, \frac{1}{\calA_0\, (1 + \scrK)\,\theta_{00}\,
    \phi^{\nicefrac{1}{\scrK}}}\, \left|\bfnabla\! \times \!
    \left(\frac{\theta_0}{\phi}\right)\!\bfiv\,\calQdot\right| 
    \nonumber \\
    & & -\, \frac{1}{\scrK\calA_0^{\,\prime}\theta_{00}}
    \left(\frac{q}{\phi}\right)\calQdot\, ,
\end{eqnarray}
where $\,q(\bfix,t) = \eta / \phi^{\nicefrac{1}{\scrK}}\,$ is the 
equivalent-barotropic potential vorticity (EBPV) with
$\eta(\bfix,t) = |\bfeta\,| \equiv |\,\bfzeta + \bfif|$ the absolute 
vorticity and $\zeta(\bfix,t) = |\bfnabla \times \bfiv|$ the 
relative vorticity.
The terms on the right hand side of Equation~(\ref{eqn:q}) represent 
the EBPV sources and sinks, by which EBPV is created and destroyed 
through gradients of $\theta$, dissipation, and heating and cooling 
at the bottom bounding surface.
Note that the EBPV manifestly---and correctly---couples barotropic 
dynamics and thermodynamics in a 2D system, unlike in the SWEs.
Note also that, under the adiabatic condition (i.e., 
$\bfCalD_{\bfiv} = \calQdot = 0$ and $\theta_0$ is a constant), the 
EBPV is materially conserved and serves as a tracer of the flow. 

Isentropic maps of (Ertel and quasi-geostrophic) potential vorticity 
have been effectively used in Earth's atmosphere studies and have 
led to great advancements \citep[see, e.g.,][]{Hosketal85}.
Similar advancements can be achieved by accurately capturing the 
EBPV using high-resolution simulations \citep{Boyd00} employing 
sophisticated methods, such as the pseudospectral method with 
high-order hyperviscosity 
\citep{Orsz70,Eliasenetal70,Choetal03,SkinCho21} and contour 
dynamics with surgery \citep{Drit88}.
Other advantages of potential vorticity is that it allows the 
effects of diabatic forcing to be accurately represented and 
isolated and the flow to be ``balanced'', to mitigate or delay the 
onset of gravity wave generation 
\citep[e.g.,][]{Fordetal00,MoheDrit01,Choetal03}.

\section{Thermal Relaxation}\label{sec:relaxation}

On CEGPs, thermal forcing may be important in the $\sim$$10^3$~Pa 
to $\sim$$10^6$~Pa region, where the radiative time scales are 
short (smaller than few planetary rotation periods). 
By driving large-scale flows away from a zonally symmetric state, 
the zonally {\it a}symmetric radiative forcing strongly enhances 
the already asymmetrizing influence of vortices and waves on CEGPs.
In some circumstances, $\calQdot$ in Equations~(\ref{eqn:debe}) 
may be represented by the ``Newtonian cooling'' approximation 
\citep[e.g.,][]{Andrewsetal87,Salby96}.
This approximation is a simple, pragmatic representation of 
radiative heating and cooling effects on the large-scale 
dynamics---a relaxation of the temperature field to a specified 
``equilibrium'' field.
In actuality, the equilibrium field depends in a complicated way 
on the mixing ratios of radiatively-active species and their 
ever-changing 3D distributions.
At present, the mixing ratio distributions and the equilibrium 
field are not known for CEGPs. 

More fundamentally, the following simplifying assumptions are 
made in the approximation that should be noted: 
1)~the vertical motion is ignored; 
2)~the magnitude of the temperature perturbations is small, 
compared to the equilibrium temperature $T_e(\bfix, \theta, t)$;
3)~the vertical gradient of the transmission function is  
$\sim$$1 / \calH_p$, where $\calH_p(\bfix, \theta, t)$ is 
the pressure scale height; and
4)~the environment is in local thermodynamic equilibrium.  
Despite these limitations, the Newtonian cooling approximation 
has been widely used in theoretical studies because of its 
practicability and because of its usefulness in appropriate 
situations or locations (e.g., near the 1~bar level on CEGPs).
Its use is consistent for tall, columnar motions---such as 
those described by the EBEs.
An important parameter in the approximation is the relaxation 
or the ``cooling'' time, 
$\tau_r(\bfix, \theta, t) \approx 
\rho c_p\, /\, (4 \sigma T_e^3\,d{\scrT} / D z^\star)$, 
where $\sigma$, $\scrT(z^\star)$, and 
$z^\star(\bfix, \theta, t) \equiv \calH_p\ln(p_{00} /p)$ are 
the Stefan--Boltzmann constant, transmission function, and 
log-pressure height, respectively. 

As in the PEs, thermal forcing enters into the EBEs through 
the thermodynamics equation, Equation~(\ref{eqn:debe}c):
\begin{eqnarray}\label{eqn:dten}
    \frac{D\theta_0}{D t} &\ =\ &
    (1 - \calA_0)\,\bfiv\!\cdot\!\bfnabla\,\theta_0\,
    -\, \alpha\left(\frac{\phi_\ast - \phi_e}{\phi}\right)\theta_0,
    \qquad
\end{eqnarray}
where 
$\calQdot = -\calA_0^{\,\prime}\,\theta_{00}(\phi_\ast - \phi_e)$.
Here, the heating/cooling is proportional to a constant relaxation 
rate $\alpha$ and the departure of $\phi_\ast$ from a radiative 
equilibrium background state $\phi_e(\bfix, t)$.
Note that the specification of the forcing in 
Equation~(\ref{eqn:dten}) is in contrast to the Newtonian cooling approximation typically implemented in SWEs studies of 
exoplanets \cite[e.g.,][]{ShowPol11}. 
In those studies, there is, of course, no thermodynamic equation 
to force in the SWEs.
Instead, a proxy is used, in which the fluid thickness (or mass) 
is relaxed to an ``equilibrium'' thickness.
In addition, since the $\phi$ and $\theta_0$ fields are coupled 
in the EBEs, Equation~(\ref{eqn:debe}b) must also be augmented 
concordantly:
\begin{eqnarray}\label{eqn:dpen}
  \frac{D \phi}{D t} &\ =\ &
        - \scrK \phi\,\bfnabla\!\cdot\!\bfiv\, +\, (1 - \calA_0)\, 
        \phi\,\bfiv\!\cdot\!\bfnabla(\ln\theta_0)\,  \qquad
        \nonumber \\
        & & +\, \alpha\,(\phi_\ast - \phi_e)\, . 
\end{eqnarray}
In Equations~(\ref{eqn:dten}) and (\ref{eqn:dpen}), note also 
the opposite sign of the forcing terms.
Cooling (i.e., when $\phi_\ast > \phi_e$) leads to loss of 
$\theta_0$ following the fluid element, as reflected in 
Equation~(\ref{eqn:dten}).

In a generic, stably stratified baroclinic environment, the loss 
of potential temperature induces a downward deflection of the 
fluid element following its motion. 
Concomitantly, the downward motion of the element leads to an 
increase in $\calH_{p_0}$; hence, the fluid column is stretched 
vertically following the motion. 
The stretching corresponds to an increase in the temperature of 
the column via the hypsometric relation \citep{Holt04,Choetal08}.
In the equivalent-barotropic formulation, this process is 
captured by an increase in $\phi$, the fluid column, following 
the motion, and is expressed by Equation~(\ref{eqn:dpen}).
The increased $\phi$ then drives the motion, which in turn 
rearranges $\theta_0$ in the presence of gradients (as well 
as from the thermal relaxation itself); 
see Equations~(\ref{eqn:debe}a) and~(\ref{eqn:dten}).
A detailed numerical investigation of the thermal forcing, as 
implemented in Equations~(\ref{eqn:dten}) and (\ref{eqn:dpen}), 
will be presented elsewhere.

We also note here that a ``negative mountain'' (i.e., 
$\langle\phi_b(\bfix, t)\rangle < 0$, where 
$\langle (\cdot) \rangle$ denotes the global-mean) here has
a similar effect on the flow, and is consistent with its use in 
\citet{Choetal03} and \citet{Choetal08}. 
In those studies, $z_0(\bfix, t) = (\calA_0 / g)\,\phi_b$ is used 
to represent the ``effects of diabatic heating'' in the context 
of adiabatic flow.
In Equation~(\ref{eqn:dten}), a negative topography formally plays 
the role of an additional, ``effective'' equilibrium and produces a corresponding decrease in $\theta_0$ following the flow when 
$|\phi_b| < (\phi - \phi_e)$; 
a $|\phi_b| \ge 0$ is not physical, as it violates the single-fluid 
assumption.

To complete the EBEs with Newtonian cooling, an 
equivalent-barotropic structure needs to be specified.
For simplicity, we choose the Lamb structure \citep{Lamb32,Bret69}:
\begin{eqnarray}
    \calA(\theta)\ =\, 
    \left[\frac{1 -\scrK}{\langle\theta_0\rangle}\right]\theta,
\end{eqnarray}
where 
$\langle\theta_0\rangle \equiv\, \langle T_0\rangle\,
(p_{00}\, /\, \langle p_0\rangle)^\scrK$.
Therefore, the bottom surface values of the structure function 
and its derivative are $\calA_0 = 1 - \scrK$ and 
$\calA_0^{\,\prime} = (\calA_0 / \langle \theta_0 \rangle)$, 
respectively.
Choosing $\scrK = 2/7$, $\langle T_0 \rangle \sim 1500$~K, and
$p_{00} = 10^6$~Pa, characteristic of the CEGP HD209458b at 
$\sim$10$^5$~Pa, we have $\calA_0 = 5/7$, 
$\langle \theta_0 \rangle \approx 2900$~K, and 
$\calA_0^{\,\prime} \approx 2.5 \times 10^{-4}$\,K$^{-1}$.
This leads to 
$\alpha$[s$^{-1}$] $\sim \calQdot / (250~{\rm K})$ for a flat 
bounding surface, giving a cooling time of  
$\ \lesssim 1$~planetary rotation period for 
$\calQdot \sim 10^{-3}$~K~s$^{-1}$.
These values are consistent with those used in current 
simulations \citep[e.g.][]{Choetal21,SkinCho22}.
For giant exoplanets which are cooler and hotter than 
HD209458b---e.g., WASP-11b \citep{Peppetal17} and 
WASP-121b \citep{Evanetal17}, 
respectively---$\langle \theta_0 \rangle$ is correspondingly 
smaller and larger than that for HD209458b. 
Hence, $\alpha$ is correspondingly smaller and larger, and 
the cooling time is also then correspondingly longer and 
shorter.

Including both Newtonian cooling and momentum dissipation, 
the Equations~(\ref{eqn:debe}) for an atmosphere with a 
Lamb structure read: 
\begin{subequations}\label{eqn:eln}
    \begin{eqnarray}
        \frac{D\bfiv}{D t}\ &\ =\ &
        - \bfnabla\phi_\ast - \bfif\!\times\!\bfiv 
        + \phi\,\bfnabla(\ln\theta_0) \nonumber \\ 
        & & +\ \alpha\left[ \frac{\phi_\ast - \phi_e}{(1 + \scrK) \,\phi}\right]\!\bfiv\, +\, \bfCalD_{\bfiv} \\
        \frac{D \phi}{D t}\ &\ =\ &
        -\scrK \phi \Big[\bfnabla\!\cdot\!\bfiv\, 
        -\, \bfiv\!\cdot\!\bfnabla(\ln\theta_0)\Big] 
        \qquad \quad \nonumber \\ 
        & & +\ \alpha(\phi_\ast - \phi_e) \\
        \frac{D\theta_0}{D t\ } &\ =\ &
        \scrK\bfiv\!\cdot\!\bfnabla\theta_0 \nonumber \\
        & & -\ \alpha\Big(\frac{\phi_\ast - \phi_e}{\phi}\Big)\theta_0\,
        +\, {\cal D}_{\theta_0}, 
    \end{eqnarray}
\end{subequations}
where a term for small-scale thermal diffusion has been quietly 
reintroduced---mainly for numerical purposes.
In solving Equations~(\ref{eqn:eln}) numerically, one may choose 
to employ hyperdiffusivities for 
$\{\bfCalD_{\bfiv}, {\cal D}_{\theta_0}\}$, to ensure momentum and 
heat flux out of the simulation only near the truncation scale 
while concurrently preventing numerical instability
\cite[e.g.,][]{ChoPol96a,ThraCho11,SkinCho21}.
In analytical work, $\bfCalD_{\bfiv}$ and ${\cal D}_{\theta_0}$
are generally set to 0, self-consistent with the focus on the 
large scales.
If $\scrK = \calA_0 = 1$, $\bfCalD_{\bfiv} = \calQdot\! = 0$, 
and $\theta_0$ is a constant, Equations~(\ref{eqn:debe}) are 
formally identical to the inviscid SWEs with bottom 
topography---plus a passive tracer. 
From this viewpoint, $\scrK$ (which is always $< 1$ for physical 
fluids) leads to an enhanced advection, or reduced divergence 
(i.e., lateral compressibility). 
Equations~(\ref{eqn:eln}) then give the following for the EBPV:
\begin{eqnarray}\label{eqn:qeln}
    \frac{D q}{D t} &\ =\ & \
    \frac{1}{\phi^{\nicefrac{1}{\scrK}}}
    \Big|\bfnabla\phi\! \times\! \bfnabla(\ln\theta_0)\Big|\, 
    -\, q\,\bfiv\!\cdot\!\bfnabla(\ln \theta_0) \nonumber \\ 
    & &\ +\  \frac{1}{\phi^{\nicefrac{1}{\scrK}}} 
    \Big|\bfnabla\!\times\! \bfCalD_{\bfiv}\Big| \nonumber \\
    & &\ +\ \alpha\,\frac{1}{(1 + \scrK)\,
    \phi^{\nicefrac{1}{\scrK}}}\, 
    \left|\bfnabla\! \times\! 
    \left(\frac{\phi_\ast - \phi_e}{\phi}\right)\!\bfiv \right| 
    \nonumber \qquad \\
    & &\ -\ \alpha\,\frac{q}{\scrK} 
    \left(\frac{\phi_\ast - \phi_e}{ \phi}\right).
\end{eqnarray}
As before, the adiabatic result is recovered when 
$\alpha = \bfCalD_{\bfiv}\, =\, 0$ and $\theta_0$ is a constant.

\section{Vorticity--Divergence Formulation}\label{sec:formulation}

The EBEs provide a useful framework for the analysis of the dynamics 
of large-scale atmospheric flow as well as a check on new numerical 
algorithms and models for solving the PEs. 
This is because many of the mathematical and computational 
properties of the PEs are embodied in the simpler EBEs. 
Making use of the full equation of state, the EBEs are the proper 
one-layer (or one-level) reduction of the PEs with lateral 
inhomogeneity in the thermodynamic variable.
In contrast, the SWEs are formally valid only for a homogeneous 
liquid.
It is also important to point out that, with free-slip boundary 
conditions at the top and bottom of the domain, the PEs do not 
admit supersonic flow \citep{Choetal15,Choetal19}, and the 
EBEs inherit this feature.
The SWEs also do not admit supersonic flow because the 3D flow, 
from which the fluid thickness (mass) equation is derived, is 
assumed to be solenoidal (i.e., nondivergent) from the 
outset \citep[see, e.g.,][]{Pedl87}.

It is useful to express Equations~(\ref{eqn:debe}), the diabatic 
EBEs, in the vorticity--divergence form for both numerical and 
analytical work.
In this form the equations read:
\begin{subequations}\label{eqn:ebvde}
    \begin{eqnarray}
        \frac{D\eta}{D t}\ &\ =\ & 
        -\, \eta\,\delta\, +\, \Big|\bfnabla \times 
        \phi\,\bfnabla(\ln\theta_0)\Big| \nonumber \\ 
        & & +\ \frac{1}{\calA_0\,(1 + \scrK)\,\theta_{00}}
        \left|\bfnabla \times \left(\frac{\theta_0}{\phi}\right)\!
        \bfiv \calQdot\right| \nonumber \\
        & & +\ \Big|\bfnabla\!\times\!\bfCalD_{\bfiv}\Big| \\
        \frac{D\delta}{D t}\ &\ =\ &
        \bfiv\!\cdot\!\bfnabla\,\delta\, 
        -\, \bfnabla\!\cdot (\bfeta\times\bfiv)
        \nonumber \\
        & & -\ \bfnabla^2\Big(\onehalf\bfiv^2 + \phi_\ast\Big)\,
        +\, \bfnabla\!\cdot\!\Big[\phi\,\bfnabla(\ln\theta_0)\Big] 
        \qquad \quad \nonumber \\
        & & +\ \frac{1}{\calA_0\,(1 + \scrK)\,\theta_{00}}\,
        \bfnabla\!\cdot\!\left[\left(\frac{\theta_0}{\phi}\right)\!
        \bfiv \calQdot\right] \nonumber \\
        & & +\ \bfnabla \cdot \bfCalD_{\bfiv} \\ 
        \frac{D \phi}{D t}\, &\ =\ &
        -\scrK \phi\,\delta\, +\, (1 - \calA_0)\, 
        \phi \bfiv\!\cdot\!\bfnabla(\ln\theta_0) \nonumber \\
        & & +\, 
        \left(\frac{1}{\calA_0^{\,\prime}\,\theta_{00}}\right)\calQdot  \\
        \frac{D\theta_0}{D t\ }\! &\ =\ &
        (1 - \calA_0)\,\bfiv\!\cdot\!\bfnabla\theta_0\, +\, \left(\frac{1}{\calA_0^{\,\prime}\,\theta_{00}\,\phi}\right)
        \theta_0\,\calQdot\, , 
    \end{eqnarray}
\end{subequations}
where $\delta(\bfix, t) \equiv \bfnabla \cdot \bfiv$ is the velocity 
divergence and $|(\cdot)|$ denotes the magnitude in the vertical 
direction.
Related forms---based on potential vorticity, for example---are also 
useful \citep[e.g.,][]{MoheDrit01,ViudDrit04} and will be considered 
in a future work.
In the latter form, hierarchies of balance conditions relating the 
divergence and {\it a}geostrophic vorticity, $f\zeta - \nabla^2\phi$, 
to the linearized potential vorticity, 
$q - f(\phi / \langle \phi \rangle)$, can be introduced.
For exoplanet studies, hierarchical balance conditions can greatly 
improve accuracy over simpler conditions (e.g., 
quasi-geostrophy)---especially at the large scales.
With Newtonian cooling and dissipations, Equations~(\ref{eqn:ebvde}) 
read: 
\begin{subequations}\label{eqn:ebvdn}
    \begin{eqnarray}
        \frac{D\eta}{D t}\ &\ =\ & 
        - \eta\,\delta\, +\, \Big|\bfnabla \times
        \phi\,\bfnabla(\ln\theta_0)\Big| \nonumber \\ 
        & & +\ 
        \frac{\alpha}{(1 + \scrK)\,\langle \theta_0 \rangle}
        \left|\bfnabla \times 
        \left(\frac{\phi_\ast - \phi_e} {\phi} \right)
        \theta_0\,\bfiv\right| \nonumber \\
        & & +\ \Big|\bfnabla\times\bfCalD_{\bfiv}\Big| \\
        \frac{D\delta}{D t}\ &\ =\ &
        \bfiv\!\cdot\!\bfnabla\,\delta\, 
        -\, \bfnabla\!\cdot (\bfeta\times\bfiv)  \nonumber \\
        & & -\ \bfnabla^2\Big(\onehalf\bfiv^2 + \phi_\ast\Big)\
        +\, \bfnabla\!\cdot\!\Big[\phi\,\bfnabla(\ln\theta_0)\Big] 
        \qquad \quad \nonumber \\
        & & +\ 
        \frac{\alpha}{(1 + \scrK)\,\langle \theta_0 \rangle}
        \bfnabla \cdot 
        \left[\left(\frac{\phi_\ast - \phi_e}{\phi}\right)
        \theta_0\,\bfiv\right] \nonumber \qquad \quad \\
        & & + \bfnabla \cdot \bfCalD_{\bfiv} \\ 
        \frac{D \phi}{D t}\, &\ =\ & -\scrK \phi\,\delta\ +\, 
        (1 - \calA_0)\, 
        \phi \bfiv\!\cdot\!\bfnabla(\ln\theta_0)\, \nonumber \\
        & & +\ \alpha(\phi_\ast - \phi_e) \\
        \frac{D\theta_0}{D t\ }\! &\ =\ &
        (1 - \calA_0)\bfiv\!\cdot\!\bfnabla\theta_0 - 
        \alpha\Big(\frac{\phi_\ast - \phi_e}{\phi}\Big)\theta_0 
        + {\cal D}_{\theta_0}. 
    \end{eqnarray}
\end{subequations}

For global simulations, a rewrite of Equations~(\ref{eqn:ebvdn})
in spherical geometry is useful; 
hence, $\bfix \rightarrow (\lambda, \varphi)$ with $\lambda$ the 
longitude and $\varphi$ the latitude.
Here in anticipation of numerical utility,\footnote{for example, 
semi-implicit time stepping \citep{Robe69,StanCote91}} we make use 
of the decompositions,
$\phi(\bfix, t) = \langle\phi\rangle + \Phi(\bfix, t)$ and 
$\theta_0(\bfix, t) = \langle\theta_0\rangle + \Theta(\bfix, t)$,
and the mapping, $\bfiv\cos\varphi\, \mapsto\, \bfiV = (U, V)$  
\citep{Robert66}:  
\begin{subequations}\label{eqn:ebspn}
    \begin{eqnarray}
        \frac{\del\eta}{\del t}\ &\ =\ &
        - \frac{1}{R_p\,(1 - \mu^2)} \frac{\del\ }{\del\lambda}
        \Big(\eta U - \alpha\varepsilon V\Big) \nonumber \\
        & & -\ \frac{1}{R_p}\frac{\del\ }{\del\mu}
        \Big(\eta V + \alpha\varepsilon U\Big) \nonumber \\
        & & +\ 
        \Big|\bfnabla \Phi\! \times\! \bfnabla \Upsilon\Big| 
        + \Big|\bfnabla\!\times\!\bfCalD_{\bfiv}\Big| \\
        \frac{\del\delta}{\del t}\ &\ =\ &
        \frac{1}{R_p\,(1 - \mu^2)}\frac{\del\ }{\del\lambda}
        \Big(\eta V + \alpha\varepsilon U\Big) \nonumber \\
        & & -\ \frac{1}{R_p}\frac{\del\ }{\del\mu}
        \Big(U\eta - \alpha\varepsilon V\Big) \nonumber \\
        & & -\ \bfnabla^2\!\left[\frac{U^2 + V^2}{2\,(1 - \mu^2)} 
        + \Big(\Phi + \phi_b \Big) 
        - \langle\phi\rangle\Upsilon \right]  \nonumber \\
        & & +\ 
        \bfnabla\!\cdot\!\Big(\Phi\,\bfnabla\,\Upsilon\Big)
        +\ \bfnabla\!\cdot\! \bfCalD_{\bfiv} \\
        \frac{\del\Phi}{\del t} & = &
        -\, \frac{1}{R_p\,(1 - \mu^2)}\frac{\del\ }{\del\lambda}\Big(U\Phi\Big)\,
        -\, \frac{1}{R_p}\frac{\del\ }{\del\mu}
        \Big(V\Phi\Big)\nonumber \\
        & & +\ \Big[(1 - \scrK)\Phi 
        - \scrK\langle\phi\rangle\Big]\delta\, 
        +\, \alpha(\phi_\ast - \phi_e) \qquad \quad \\
        \frac{\del\Theta}{\del t} & = &
        -\, \frac{\calA_0}{R_p\,(1 - \mu^2)}
        \frac{\del\ }{\del\lambda}\Big(U\Theta\Big)\,
        -\, \frac{\calA_0}{R_p}
        \frac{\del\ }{\del\mu}\Big(V\Theta\Big)\, \nonumber \\
        & & -\ \calA_0\,\Theta\,\delta\, 
        -\ \alpha\Big(\frac{\phi_\ast - \phi_e}{\phi}\Big)
        \theta_0\, 
        +\, {\cal D}_{\theta_0}.  
    \end{eqnarray}
\end{subequations}
Here
\begin{eqnarray}
    \Upsilon(\bfix, t) &\ =\ &\,
    \ln\!\left(1 + \frac{\Theta}{\langle \phi_0 \rangle}\right) \\
    \varepsilon(\bfix, t) &\ =\ &\! \frac{1}{(1 + \scrK)}
    \left(\frac{\phi_\ast - \phi_e}{\phi}\right)\!\!
    \left(\frac{\theta_0}{\langle \theta_0 \rangle}\right) 
    \quad \quad
\end{eqnarray}
and
\begin{subequations}\label{eqn:vdspn}
    \begin{eqnarray}
        \eta &\ =\ &
        \frac{1}{R_p\,(1 - \mu^2)}\frac{\del V}{\del\lambda}\,
        -\, \frac{1}{R_p}\frac{\del U}{\del\mu}\, +\, f \qquad \\
        \delta &\ =\ &
        \frac{1}{R_p\,(1 - \mu^2)}\frac{\del U}{\del\lambda}\,
        +\, \frac{1}{R_p}\frac{\del V}{\del\mu},
    \end{eqnarray}
\end{subequations}
where $R_p$ is the planetary radius, and $\mu = \sin\varphi$.
In Equations~(\ref{eqn:ebspn}), we have left some terms (e.g., 
involving quadratic product of $\Phi$ and $\Upsilon$ derivatives 
and $\bfCalD_{\bfiv}$) in coordinate-free form for notational 
expediency; 
all the dissipation terms ($\bfCalD_{\bfiv}$ and 
${\cal D}_{\theta_0}$) are, in fact, likely to be replaced in 
numerical simulation work---e.g., with hyperdissipation 
\citep{Polietal14,SkinCho21}.
We remind the reader that in spherical coordinates
\begin{eqnarray*}
    \bfnabla\,\xi &\ =\ & \left(\frac{1}{R_p \sqrt{1 - \mu^2}}
    \frac{\del \xi }{\del \lambda},\  
    \frac{\sqrt{1 - \mu^2}}{R_p}
    \frac{\del \xi }{\del \mu} \right)  \\
    \bfnabla\!\cdot\!\bfF &\ =\ & \frac{1}{R_p \sqrt{1 - \mu^2}}
    \frac{\del F_\lambda}{\del \lambda}\, + \frac{1}{R_p}
    \frac{\del\ }{\del \mu}\Big(F_\varphi \sqrt{1 - \mu^2}\Big)   \\
    |\bfnabla\! \times\! \bfF| &\ =\ &  
    \frac{1}{R_p \sqrt{1 - \mu^2}} 
    \frac{\del F_\varphi }{\del \lambda} - \frac{1}{R_p}
    \frac{\del\ }{\del \mu}\Big(F_\lambda \sqrt{1 - \mu^2}\Big) 
    \quad \\
    \bfnabla^2 \xi &\ =\ & \frac{1}{R_p^2\,(1 - \mu^2)}
    \frac{\del^2 \xi }{\del \lambda^2}\, + \frac{1}{R_p^2}
    \frac{\del^2\ }{\del \mu^2}\Big[\xi(1 - \mu^2)\Big],
\end{eqnarray*}
where $\xi(\lambda, \mu, t)$ and
$\bfF(\lambda, \mu, t) = (F_\lambda, F_\varphi)$
are arbitrary scalar and vector fields, respectively. 

Using Helmholtz--Hodge decomposition, the $\bfiV$ field can be 
split into two parts: 
$\bfiV = \bfnabla\chi - \bfnabla\!\times\!\bfpsi$, where 
$\chi(\bfix, t)$ is the velocity potential and $\bfpsi$ is
a vector in the vertical direction such that
$|\bfpsi| = \psi(\bfix, t)$ is the streamfunction.
This gives
\begin{subequations}\label{eqn:UV}
    \begin{eqnarray}
        U &\ =\ &
        \frac{1}{R_p}\frac{\del\chi}{\del\lambda}\,
        -\, \frac{1 - \mu^2}{R_p}\frac{\del\psi}{\del\mu} \\
        V &\ =\ &
        \frac{1}{R_p}\frac{\del\psi}{\del\lambda}\,
        +\, \frac{1 - \mu^2}{R_p}\frac{\del\chi}{\del\mu}\, .
    \end{eqnarray}
\end{subequations}
Therefore, $\eta\ =\ \nabla^2\psi + f$ and 
$\delta\ =\ \nabla^2\chi$.
Then, all of the scalar field variables can be conveniently 
represented in spherical harmonic series \citep{Mech79}; 
for example, 
\begin{eqnarray}
    \xi(\lambda,\mu,t) &\ =\ & 
    \sum_{n = |m|}^N\,\sum_{m = -M}^{N(m)} 
    \hat{\xi}_n^m(t)\, {\cal P}_n^m(\mu)\, e^{im\lambda}, \qquad
\end{eqnarray}
where $(M,N)$ are the truncation wavenumbers and a general 
pentagonal truncation representation is presented;
when $M = N$, we have the usual triangular truncation 
\cite[e.g.,][]{Polietal14,SkinCho21}.
The spectral method presents a natural solution to the problems 
in spherical geometry \citep{Boyd00}. 

\section{Discussion}\label{sec:discussion}

In this letter, we have presented the diabatic EBEs, which 
provide a useful framework for the analysis of the dynamics of 
large-scale atmospheric flows on CEGPs, as well as the analysis 
of innovative numerical methods that might be applied to the 
solutions of the EBEs and PEs.  
The EBEs are easily adaptable to the spectral transform method, 
especially for the spherical geometry.
Understanding the response of exoplanet atmospheres to both 
radiative and mechanical forcing accurately is very challenging. 
Modeling the atmospheres requires a range of approaches, from 
simple analytical calculations to full numerical simulations 
of the general circulation and climate.
Here we have presented equation sets which represent a proper 
2D reduction of the 3D dynamics.
The reduced equations correctly describe the dynamics of CEGP 
atmospheres in a single-layer context.
The equations also apply to any atmospheres in which the 
motions are columnar over a depth of few scale heights.

Robust properties that have slowly emerged from careful 
simulations over many years motivate this letter.
First and foremost, high resolution simulations of most CEGPs 
exhibit a strong barotropicity over $\sim$2 pressure scale 
heights \citep[e.g.,][]{SkinCho22,Skinetal23}.
Given that computational resources are still prohibitive for 
well-resolved 3D simulations over the full stably stratified 
region \citep{SkinCho21,SkinCho24}, we believe that exoplanet 
characterization studies would benefit greatly from more 
detailed analyses and simulations using the EBEs.
Second, the strong rotational influence on the flow is now 
much more evident---regardless of the amplitude and 
non-migration of the thermal forcing.
The most important influence is on the the establishment of 
a strong {\it azonal} character: 
large vortices are generated in both the equatorial and polar 
regions.
Here, given that close-in planets generally possess a large 
deformation scale, the aforementioned two regions---as well as 
the day and night sides, if 1:1 spin--orbit synchronized---must 
influence each other strongly, as have been pointed out in 
\citet{Choetal03} and \citet{Choetal08}. 
Finally, the adiabatic EBEs appear to be able capture the full 
range of global temperature distributions on CEGPs---including 
rotating, oscillating, shifted, and fixed day--night 
distributions. 
A more realistic assessment with diabatic EBEs would be useful,
given the less ad hoc nature of the equations compared with the 
forced SWEs \citep[see, e.g.,][]{Choetal08}.

The strongly barotropic and extremely wide scale-range 
characters of the numerical solutions of the diabatically forced 
PEs suggest focusing on the (columnar) lateral dynamics at 
resolutions greater than currently achieved is useful.
This is sensible because, despite the vertical integration, 
the EBEs support many of the fluid motion supported by the 
PEs---includeing Rossby waves, gravity waves, balanced motions 
(e.g., geostrophic), adjustments, and barotropic instability. 
Some of the consequences of baroclinic processes, such as stirring 
by eddies or convection, may also be instructively represented 
and studied in the aptly reduced model. 
The high resolution permitted also allows mixing of potential 
vorticity and other fine-scale tracers (e.g., CO$_2$, H$_2$O, 
and clouds) to be captured down to the small scales.
Moreover, the EBEs have a consistent set of conservation 
laws for ``mass'', energy, angular momentum, potential vorticity, 
potential enstrophy, as well as more ``exotic'' quantities such 
as pseudomomentum \citep[see, e.g.,][]{Choetal08} that also 
advance our understanding of exoplanet atmospheres. 

\section*{Acknowledgments}
The authors thank Jack W. Skinner and Quentin Changeat for helpful 
discussions.

\bibliography{references}
\bibliographystyle{aasjournal}

\end{document}